\documentclass[twocolumn,showpacs,preprintnumbers,amsmath,amssymb]{revtex4}

\usepackage{graphicx}
\usepackage{dcolumn}
\usepackage{bm}
\usepackage{multirow}
\usepackage{hyperref}

\def\e{\mathrm{e}}
\def\R{{\mathbb{R}}}

\begin{document}

\title{Tuning symplectic integrators is easy and worthwhile}

\author{Robert I McLachlan}
\email{r.mclachlan@massey.ac.nz}
\affiliation{School of Fundamental Sciences, Massey University, New Zealand}
\date{\today}

\begin{abstract}
Many applications in computational physics that use numerical integrators based on splitting and composition can benefit from the development of optimized algorithms and from choosing the best ordering of terms. The cost in programming and execution time is minimal, while the performance improvements can be large.
\end{abstract}

\pacs{Valid PACS appear here}
\maketitle

\section{\label{sec:level1}Introduction}

Symplectic numerical integration by splitting 
the Hamiltonian and composing the flows of the associated vector fields has become
an extremely widely used technique in computational science, especially computational physics and 
chemistry. A splitting into two parts, $H=H_1+H_2$, together with the 3-term composition
$$\e^{\frac{1}{2}h X_1}\e^{h X_2} \e^{\frac{1}{2}h X_1},$$
is the most common and is often all that is needed. It is variously called the leapfrog, St\"ormer--Verlet, or Strang splitting method. (Here $X_i$ is the Hamiltonian vector field associated with Hamiltonian $H_i$, $\e^{h X}$ is the time-$h$ flow of the vector field $X$, and $h$ is the time step.) For example, it is the basic building block of the Hamiltonian Monte Carlo method.

Some Hamiltonians can only be written as the sum of more than two explicitly integrable terms, say $H=\sum_{i=1}^n H_i$. In addition, some applications need order higher than two to achieve the required accuracy for a given amount of computational effort. Methods of all orders exist, but are progressively more expensive. Optimized methods have been develop that can significantly reduce discretization errors at fixed cost \cite{blanes2008splitting}.

However, an informal survey of the current literature suggests that unoptimized composition methods of order 4, 6, and 8 \cite{yoshida1990construction} are in common use in cosmology, celestial mechanics, quantum mechanics, quantum statistical mechanics, solid state physics, kinetic theory, plasma physics, molecular dynamics, optics, neural networks, and fluid mechanics \cite{biondini2020semiclassical,bravetti2020numerical,budd2021classification,cai2020efficient,cho2021conservative,deng2020use,faehrmann2021randomizing,faver2020solitary,figueroa2020art,klein2020numerical,lozanov2020gfire,mancini2020computational,manwadkar2020chaos,ohno2020,seki2021quantum,sheng2020fast,song2020generating,stern2021self,tong2021symplectic,wang2020slow,wang2020wave,wang2021construction,wang2021construction3,wu2021first,xiao2021explicit}. Computations in these fields could benefit from experience gained in numerical analysis to reduce errors and error growth at little cost either in programming or execution time.

In this note we review four such techniques.

\section{Optimized methods for multi-term splittings}
Let the vector field $X=\sum_{i=1}^n$ be split into $n$ parts, each explicitly integrable and let
$$\chi_h = \e^{h X_n} \circ\dots\circ \e^{h X_1}$$
be a first-order integrator for $X$. We define the adjoint $\chi^*_h$ of $\chi_h$ by
$\chi^*_h = (\chi^{-1})_{-h}$; that is,
$$\chi^*_h = \e^{h X_1} \circ\dots\circ \e^{h X_n}.$$
Let
\begin{equation}
\label{eq:comp}
\psi_h(\alpha)= \chi_{\alpha_s h} \circ \chi^*_{\alpha_{(s-1)}h} \circ\dots\circ \chi_{\alpha_2 h} \circ\chi^*_{\alpha_1 h}.
\end{equation}
The second-order leapfrog method is $S_{2,h} = \psi_h(\frac{1}{2},\frac{1}{2})$.
Note that when $n=2$, this reduces to a composition that alternates steps of $X_1$ and $X_2$. Methods of this type designed for $n=2$ and for arbitrary $n$ have the same order \cite{mclachlan1995numerical}, although the optimal coefficients may not be the same. Nevertheless, coefficients optimised for the method \eqref{eq:comp}  can reduce the error significantly. 

In particular, the minimal-$s$ methods formed by recursively increasing the order from $k$ to $k+2$ 
by $S_{k,c h}\circ S_{k,(1-2c)h} \circ S_{k,ch}$ with $c=1/(2-2^{1/(2k+1)})$ have very large error constants and poor stability and should be avoided. The 3-stage method with $k=2$, called S$_3$4, will be used as a reference method here. (The notation indicates that it has order 4 and uses work equivalent to 3 leapfrog steps.) Its large substeps, $1.35h$, $-1.70h$, and $1.35 h$, contribute to its large error constants and poor stability. 

The optimized method that we will use for the purpose of illustration here is called BM$_6 4$ and was introduced by Blanes and Moan \cite{blanes2002practical}. It has $s=12$ and
$\alpha_1=\alpha_{12} = 0.0792036964311957$,
$\alpha_3=\alpha_{10}=     0.2228614958676077$,
$\alpha_5=\alpha_{8}=     0.3246481886897062$,
$\alpha_7=\alpha_{6}=     0.1096884778767498$,
$\alpha_9=\alpha_{4}=   -0.3667132690474257$, and
$\alpha_{11}=\alpha_{2}=     0.1303114101821663$.
BM$_6 4$ has errors are around 500 times smaller than those of S$_34$ and also allows the use of larger time steps.

\section{Order matters}
The ordering of the $n$ terms affects the errors. It can also affect the cost slightly, as there is an opportunity to collect two steps of  $X_1$  and of $X_n$ into single steps. There is no easy way to anticipate the best ordering. Fasso \cite{fasso2003comparison} studied this question in detail for the splitting of the rigid body Hamiltonian into three terms, finding that the ordering affected the error by a factor of 10, 100, or more, and that the best ordering depended in a non-obvious way on the moments of inertia of the body.

 If $n$ is not too large, one approach is to try all permutations of the $n$ terms with a typical initial condition and choose the ordering with the smallest local truncation error.

In Table \ref{tab:order} we give the results of numerical experiments testing the influence of the term ordering on the error of the optimized 4th order integrator BM$_64$. Linear systems in $\mathbb{R}^{16}$ with random coefficients are integrated for time 1 with time-step $0.1$. The difference between the best and worst orders is significant and increases with $n$. The principal error in 4th order symmetric integrators contains $\mathcal{O}(n^3)$ terms, each a triple commutator of the terms of the ODE. Some of these terms might cancel each other; the experiment suggests that the effect of using the best ordering scales
like $n^c$ with $c$ in the region 1.5 to 2.5. The effect in these unstructured systems is less that that observed by Fasso \cite{fasso2003comparison} for the (more highly structured) rigid body.

\section{Compensated summation reduces round-off error}
Round-off error is not often a dominant source of error, but in long integrations it does accumulate and can sometimes dominate the truncation error. It can be drastically reduced at minuscule cost (both in programming and execution) by the technique of compensated summation \cite{higham2002accuracy}. In each update $x_{n+1} = x_n + h y_n$, the increment $h y_n$ is computed to full precision, but $x_n$ is stored to higher precision by storing a pair of full precision numbers. Compensated summation reduces roundoff errors by a factor $\|h y_n\|/\|x_n\|$. 

Computing $h y_n$ to full precision is important and may need attention, especially if
the flow $\e^{h X}$ involves special functions. One non-trivial example that we will use below involves
$H = \frac{1}{2} q^n p^2$, an instance of Channell's observation that any monomial Hamiltonian is explicitly integrable in terms of elementary functions \cite{channell1986explicit}. First, the solution of Hamilton's equations should be presented algebraically with the update separated out:
\begin{align*}
\dot q &= q^n p,\qquad q(0)=q_0\\
\dot p &= -\frac{n}{2} q^{n-1} p, \quad p(0)=p_0 \\
\Rightarrow 
q(t) &= q_0 \left(1 + \delta\right)^{\frac{2}{2-n}}\\
&= q_0 + q_0\left( (1+\delta)^{\frac{2}{2-n}}-1\right)\quad(n\ne 2)\\
p(t) &= p_0 \left(1 + \delta\right)^{\frac{n}{n-2}}\\
&= p_0 + p_0\left((1+\delta)^{\frac{n}{n-2}}-1\right),\\
\delta &= t\left(1-\frac{n}{2}\right) q_0^{n-1} p_0
\end{align*}
Second, the update should be computed to full precision. In this case this can be achieved using the functions $\mathrm{expm1}(z)=\e^z-1$ and $\mathrm{log1p}(z) = \log(1+z)$ provided in most mathematics libraries:
\begin{align}
\begin{split}
\label{eq:kin1}
q(t) &= q_0 + q_0 \mathrm{expm1}\left(\frac{2}{2-n}\mathrm{log1p}(\delta)\right)\\
p(t) &= p_0 + p_0 \mathrm{expm1}\left(\frac{n}{n-2}\mathrm{log1p}(\delta)\right)
\end{split}
\end{align}
Similarly for $n=2$, we have
\begin{align}
\begin{split}
\label{eq:kin2}
q(t) &= \e^{p_0 q_0 t} q_0 = q_0 + q_0 \mathrm{expm1}(p_0 q_0 t) \\
p(t) &= \e^{-p_0 q_0 t} p_0 = q_0 + p_0 \mathrm{expm1}(-p_0 q_0 t) 
\end{split}
\end{align}
These formulations are  faster and more accurate than the direct implementation using powers.

Secondly, ``Brouwer's Law''  states that roundoff errors should accumulate like a random walk, leading to square-root growth in time \cite{hairer2008achieving}. Linear error growth of roundoff errors, which is often observed in momenta that would be conserved exactly in exact arithmetic, is sign of a failure of Brouwer's Law and an indication of systematic bias in the floating point arithmetic. Techniques for detecting and correcting this bias are found in the literature 
\cite{antonana2018efficient,mclachlan2014symplectic,rein2015ias15,rodriguez2012reducing,seyrich2012symmetric,symes2016efficient,vilmart2008reducing}.

\begin{table}
\begin{center}
\begin{tabular}{|c|ccc|ccc|}
\hline
 & \multicolumn{3}{c|}{antisymmetric case} & \multicolumn{3}{c|}{general case}\\
$n$ & $\min R$ & $\mathop\mathrm{med} R$ & $\max R$ &$\min R$ & $\mathop\mathrm{med} R$ & $\max R$ \\
\hline
3 &  1.16&  1.70&  4.05&    1.20&   1.90&   3.95\\ 
4  & 1.66&    2.57& 5.32&    1.98&   3.40&   7.68\\
5  & 2.23&    3.46& 7.64&    2.80&   5.41&  11.39\\
6  & 3.46&    5.00 &9.71&    4.38&   9.29&  18.55\\
7  & 4.80&   6.36&15.62&     6.41&  13.35&  24.22\\
\hline
\end{tabular}
\caption{\label{tab:order}Linear systems $\dot X = (\sum_{i=1}^n A_i) X$, $0\le t\le 1$, $X\in \R^{16\times 16}$, $X(0)=I$, are integrated using an optimized 4th order splitting method. In each trial, all $n!$ orderings of $A_1,\dots,A_n$ are tested and the ratio $R$ between the greatest and the least global error of all orderings at $t=1$ is computed. The minimum, median, and maximum ratio $R$ over 50 trials is reported. On the left (`antisymmetric case'), each $A_i$ is antisymmetric with independent normally distributed entries, modelling periodic systems with global errors growing linearly in time; on the right (`general case'), each $A_i$ has independent normally distributed entries, modelling systems with sensitive dependence on initial conditions.}
\end{center}
\end{table}

\section{Methods with processing}
Methods of the form $\rho\circ\psi\circ\rho^{-1}$ are called ``processed'' or ``corrected'' methods \cite{blanes2006composition}. The basic method $\psi$ is called the kernel, and $\rho$ the processor. Processed methods were put to spectacular use by Wisdom \cite{wisdom1996symplectic}, in which the stored values of a very long solar system simulation (based on leapfrog with a 2-term splitting) were processed {\em years after the original calculation} to dramatically reduce errors. (Or to put it another way, to reveal the true, underlying errors.) Methods with processing have been optimized for many different cases of splitting: the general case (considered in this note); near-integrable systems (as in the Wisdom--Holman example); and systems based on splitting into kinetic and potential parts. The processor can either be a method of the same class as the kernel, or calculated more cheaply by other methods. Either way, if output is taken every $m$ steps, then only $\rho\circ\psi^m\circ\rho^{-1}$ is calculated. Observables that are invariant under conjugation by the diffeomorphism $\rho$ (such as invariant phase space structures, Lyapunov exponents, periods of periodic orbits) can be computed without applying $\rho$ at all. The error can now be reduced relative to the non-corrected case because only those error terms in the kernel that cannot be removed by a corrected need to be minimized.

 In the examples of linear systems in Table \ref{tab:order}, it reduce errors by a factor of about 1.4. Greater improvements have been reported at higher order, for nearly integrable systems, and for systems with Hamiltonians of the form kinetic plus potential \cite{blanes2006composition}.

\section{Example: particles near black holes}
We consider two examples of explicit symplectic integrators based on splitting and composition. Both 
have four-dimensional (reduced) space space. The
first tracks charged particles near a Schwarzschild black hole \cite{wang2021construction} and
has a Hamiltonian separable into 4 terms; the second tracks charged particles near a Reissner--Nordstr\"om  anti-de Sitter black hole \cite{wang2021construction3} and has a Hamiltonian separable into 6 terms.
In both cases the Hamiltonians are of the form $V(q) + \sum_{i=1}^{n-1} T_i(q,p)$ where each
kinetic energy term $T_i$ is a monomial in $q$ and $p$ and hence explicitly integrable in terms of elementary functions \cite{channell1986explicit} as in Eqs (\ref{eq:kin1},\ref{eq:kin2}).

The optimized method can reduce energy errors by a factor of up to $10^4$. In all cases in the best ordering of terms the potential term was applied first, but the ordering of the remaining kinetic energy terms had to be determined experimentally.

\begin{figure}
\includegraphics[width=\columnwidth]{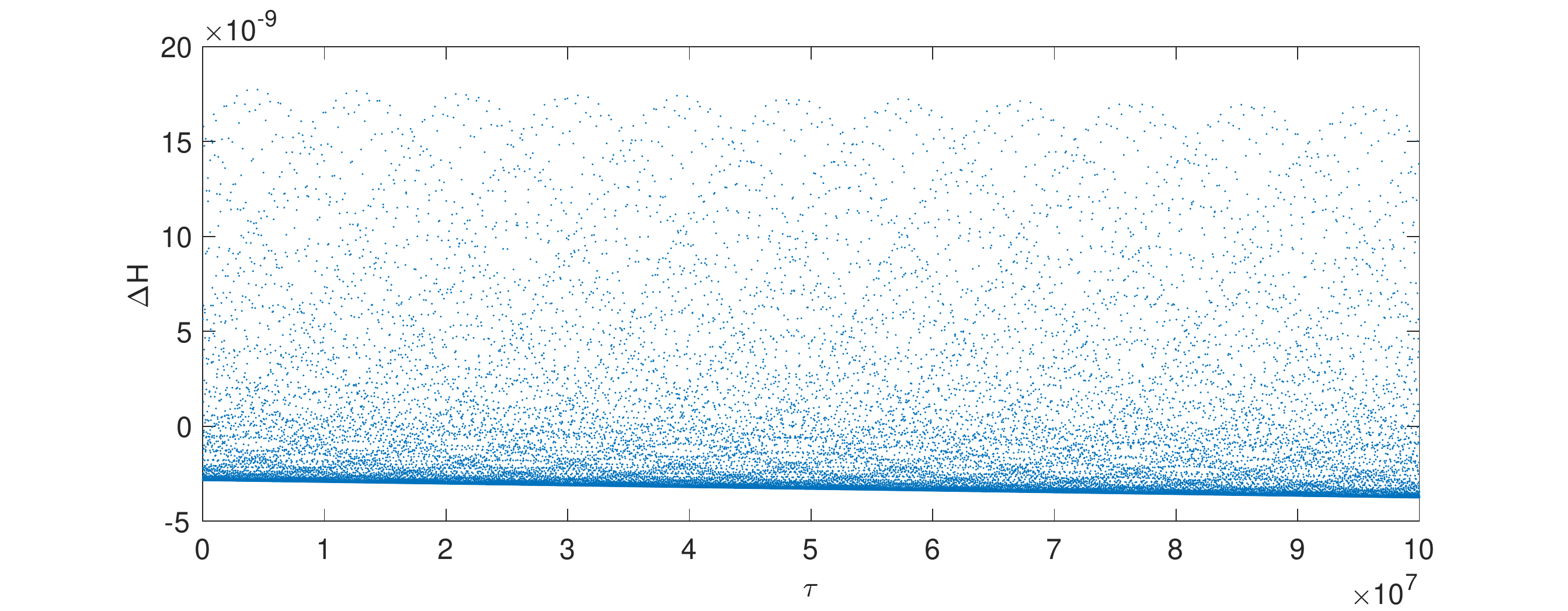}\\
\includegraphics[width=\columnwidth]{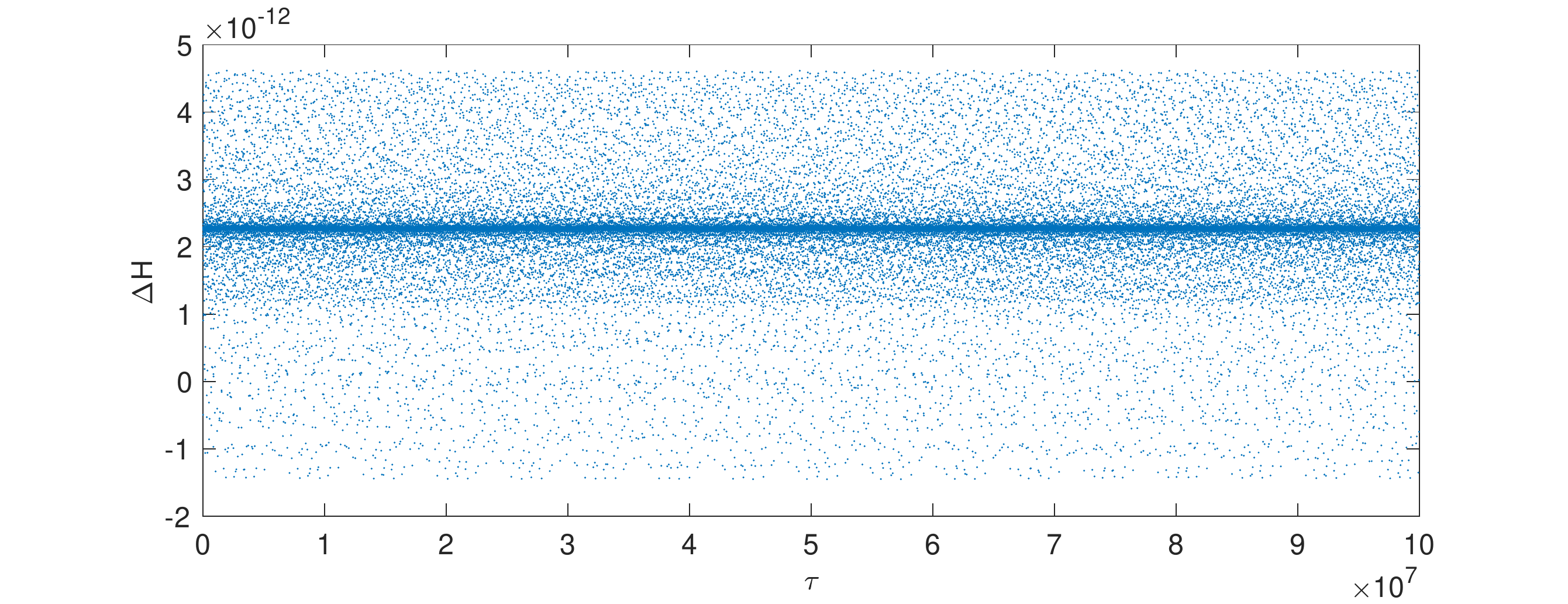}\\
\caption{\label{fig:black} Energy error in integration of a particle near a Schwarzschild black hole over proper time $\tau=10^8$ with time step $\Delta\tau=1$ by two 4th order symplectic integrators, S$_34$ (top) and BM$_64$ with compensated summation (bottom), the latter showing smaller errors and no energy drift.}
\end{figure}

In addition, the use of compensated summation drastically reduces the roundoff error to the point that it is invisible in Figure \ref{fig:black}. On closer inspection, it is not accumulating linearly, and the roundoff error after $10^8$ steps is about $2\times 10^{-16}$. In constrast, the method S$_34$ without compensated summation shows a linear drift in energy of about $10^{-9}$ over the same time interval.

\begin{table}
\begin{tabular}{|l|l|llr|}
\hline
geometry & method &  least & greatest & ratio \\
 \hline
\multirow{2}{*}{Schwarzschild} & BM$_64$ & $4.0\times 10^{-12}$ & $1.2\times 10^{-10}$ & 30 \\
& S$_34$ & $3.9\times 10^{-9}$ & $3.0\times 10^{-8}$ & 8\\
\hline
\multirow{2}{*}{R--N adS} & BM$_64$ & $1.3\times 10^{-11}$ & $3.1\times 10^{-10}$ & 24\\
& S$_34$ & $2.9\times 10^{-9}$& $6.8\times 10^{-8}$ & 23\\
\hline
\end{tabular}
\caption{\label{tab:blackhole}Energy errors in particle tracking near black holes in two cases by each of two different methods. The Schwarzschild case integrates a  quasiperiodic orbit (`Orbit 1' from \cite[Fig. 1]{wang2021construction} with 4-part splitting the R--N adS case integrates a chaotic orbit \cite[Fig. 2(c)]{wang2021construction3} with 6-part splitting. The least and the greatest energy errors over the different orderings of the terms, and their ratio, are given. The reference energy level is $-0.5$ and the energy error is measured in the sup $(\|\cdot\|_\infty$) norm; the time step is 1 in all cases.}
\end{table}

In the Schwarzschild example, the optimized 4th order processed method P$_64$ of \cite{blanes2006composition} has similar errors to BM$_64$.

\section{Discussion}
While second-order Strang splitting is sufficient for many applications, when higher accuracy is required, the optimized higher order methods are preferred. We know of no case in which the method S$_34$ is preferred over BM$_64$; the latter generally has errors orders of magnitude smaller and can be used with larger time steps. Compensated summation, and checking that roundoff errors are unbiased, costs almost nothing in programming or execution time. Despite this, the method S$_34$ without compensated summation remains widely used. 

Possibly one reason for this situation is that these algorithms are so simple and flexible, and are used in such diverse applications in computational science, that they are often implemented by hand afresh for each new application. In contrast, numerical methods that are more complicated, such as the finite element method on unstructured meshes, have passed into packages and open-source platforms that incorporate advances as they become available.

\acknowledgements The author thanks Ying Wang and Sergio Blanes for discussions.

\bibliography{4things}

\begin{thebibliography}{42}
\expandafter\ifx\csname natexlab\endcsname\relax\def\natexlab#1{#1}\fi
\expandafter\ifx\csname bibnamefont\endcsname\relax
  \def\bibnamefont#1{#1}\fi
\expandafter\ifx\csname bibfnamefont\endcsname\relax
  \def\bibfnamefont#1{#1}\fi
\expandafter\ifx\csname citenamefont\endcsname\relax
  \def\citenamefont#1{#1}\fi
\expandafter\ifx\csname url\endcsname\relax
  \def\url#1{\texttt{#1}}\fi
\expandafter\ifx\csname urlprefix\endcsname\relax\def\urlprefix{URL }\fi
\providecommand{\bibinfo}[2]{#2}
\providecommand{\eprint}[2][]{\url{#2}}

\bibitem[{\citenamefont{Blanes et~al.}(2008)\citenamefont{Blanes, Casas, and
  Murua}}]{blanes2008splitting}
\bibinfo{author}{\bibfnamefont{S.}~\bibnamefont{Blanes}},
  \bibinfo{author}{\bibfnamefont{F.}~\bibnamefont{Casas}}, \bibnamefont{and}
  \bibinfo{author}{\bibfnamefont{A.}~\bibnamefont{Murua}},
  \bibinfo{journal}{arXiv preprint arXiv:0812.0377}  (\bibinfo{year}{2008}).

\bibitem[{\citenamefont{Yoshida}(1990)}]{yoshida1990construction}
\bibinfo{author}{\bibfnamefont{H.}~\bibnamefont{Yoshida}},
  \bibinfo{journal}{Physics letters A} \textbf{\bibinfo{volume}{150}},
  \bibinfo{pages}{262} (\bibinfo{year}{1990}).

\bibitem[{\citenamefont{Biondini and
  Oregero}(2020)}]{biondini2020semiclassical}
\bibinfo{author}{\bibfnamefont{G.}~\bibnamefont{Biondini}} \bibnamefont{and}
  \bibinfo{author}{\bibfnamefont{J.}~\bibnamefont{Oregero}},
  \bibinfo{journal}{Studies in Applied Mathematics}
  \textbf{\bibinfo{volume}{145}}, \bibinfo{pages}{325} (\bibinfo{year}{2020}).

\bibitem[{\citenamefont{Bravetti et~al.}(2020)\citenamefont{Bravetti, Seri,
  Vermeeren, and Zadra}}]{bravetti2020numerical}
\bibinfo{author}{\bibfnamefont{A.}~\bibnamefont{Bravetti}},
  \bibinfo{author}{\bibfnamefont{M.}~\bibnamefont{Seri}},
  \bibinfo{author}{\bibfnamefont{M.}~\bibnamefont{Vermeeren}},
  \bibnamefont{and} \bibinfo{author}{\bibfnamefont{F.}~\bibnamefont{Zadra}},
  \bibinfo{journal}{Celestial Mechanics and Dynamical Astronomy}
  \textbf{\bibinfo{volume}{132}}, \bibinfo{pages}{1} (\bibinfo{year}{2020}).

\bibitem[{\citenamefont{Budd et~al.}(2021)\citenamefont{Budd, van Gennip, and
  Latz}}]{budd2021classification}
\bibinfo{author}{\bibfnamefont{J.}~\bibnamefont{Budd}},
  \bibinfo{author}{\bibfnamefont{Y.}~\bibnamefont{van Gennip}},
  \bibnamefont{and} \bibinfo{author}{\bibfnamefont{J.}~\bibnamefont{Latz}},
  \bibinfo{journal}{GAMM-Mitteilungen} \textbf{\bibinfo{volume}{44}},
  \bibinfo{pages}{e202100004} (\bibinfo{year}{2021}).

\bibitem[{\citenamefont{Cai and Zhang}(2020)}]{cai2020efficient}
\bibinfo{author}{\bibfnamefont{J.}~\bibnamefont{Cai}} \bibnamefont{and}
  \bibinfo{author}{\bibfnamefont{H.}~\bibnamefont{Zhang}},
  \bibinfo{journal}{Applied Mathematics Letters}
  \textbf{\bibinfo{volume}{102}}, \bibinfo{pages}{106158}
  (\bibinfo{year}{2020}).

\bibitem[{\citenamefont{Cho et~al.}(2021)\citenamefont{Cho, Boscarino, Russo,
  and Yun}}]{cho2021conservative}
\bibinfo{author}{\bibfnamefont{S.~Y.} \bibnamefont{Cho}},
  \bibinfo{author}{\bibfnamefont{S.}~\bibnamefont{Boscarino}},
  \bibinfo{author}{\bibfnamefont{G.}~\bibnamefont{Russo}}, \bibnamefont{and}
  \bibinfo{author}{\bibfnamefont{S.-B.} \bibnamefont{Yun}},
  \bibinfo{journal}{Journal of Computational Physics} p.
  \bibinfo{pages}{110281} (\bibinfo{year}{2021}).

\bibitem[{\citenamefont{Deng et~al.}(2020)\citenamefont{Deng, Wu, and
  Liang}}]{deng2020use}
\bibinfo{author}{\bibfnamefont{C.}~\bibnamefont{Deng}},
  \bibinfo{author}{\bibfnamefont{X.}~\bibnamefont{Wu}}, \bibnamefont{and}
  \bibinfo{author}{\bibfnamefont{E.}~\bibnamefont{Liang}},
  \bibinfo{journal}{Monthly Notices of the Royal Astronomical Society}
  \textbf{\bibinfo{volume}{496}}, \bibinfo{pages}{2946} (\bibinfo{year}{2020}).

\bibitem[{\citenamefont{Faehrmann et~al.}(2021)\citenamefont{Faehrmann,
  Steudtner, Kueng, Kieferova, and Eisert}}]{faehrmann2021randomizing}
\bibinfo{author}{\bibfnamefont{P.~K.} \bibnamefont{Faehrmann}},
  \bibinfo{author}{\bibfnamefont{M.}~\bibnamefont{Steudtner}},
  \bibinfo{author}{\bibfnamefont{R.}~\bibnamefont{Kueng}},
  \bibinfo{author}{\bibfnamefont{M.}~\bibnamefont{Kieferova}},
  \bibnamefont{and} \bibinfo{author}{\bibfnamefont{J.}~\bibnamefont{Eisert}},
  \bibinfo{journal}{arXiv preprint arXiv:2101.07808}  (\bibinfo{year}{2021}).

\bibitem[{\citenamefont{Faver et~al.}(2020)\citenamefont{Faver, Goodman, and
  Wright}}]{faver2020solitary}
\bibinfo{author}{\bibfnamefont{T.~E.} \bibnamefont{Faver}},
  \bibinfo{author}{\bibfnamefont{R.~H.} \bibnamefont{Goodman}},
  \bibnamefont{and} \bibinfo{author}{\bibfnamefont{J.~D.}
  \bibnamefont{Wright}}, \bibinfo{journal}{Zeitschrift f{\"u}r angewandte
  Mathematik und Physik} \textbf{\bibinfo{volume}{71}}, \bibinfo{pages}{1}
  (\bibinfo{year}{2020}).

\bibitem[{\citenamefont{Figueroa et~al.}(2020)\citenamefont{Figueroa, Florio,
  Torrenti, and Valkenburg}}]{figueroa2020art}
\bibinfo{author}{\bibfnamefont{D.~G.} \bibnamefont{Figueroa}},
  \bibinfo{author}{\bibfnamefont{A.}~\bibnamefont{Florio}},
  \bibinfo{author}{\bibfnamefont{F.}~\bibnamefont{Torrenti}}, \bibnamefont{and}
  \bibinfo{author}{\bibfnamefont{W.}~\bibnamefont{Valkenburg}},
  \bibinfo{journal}{arXiv preprint arXiv:2006.15122}  (\bibinfo{year}{2020}).

\bibitem[{\citenamefont{Klein and Stoilov}(2020)}]{klein2020numerical}
\bibinfo{author}{\bibfnamefont{C.}~\bibnamefont{Klein}} \bibnamefont{and}
  \bibinfo{author}{\bibfnamefont{N.}~\bibnamefont{Stoilov}},
  \bibinfo{journal}{Studies in Applied Mathematics}
  \textbf{\bibinfo{volume}{145}}, \bibinfo{pages}{36} (\bibinfo{year}{2020}).

\bibitem[{\citenamefont{Lozanov and Amin}(2020)}]{lozanov2020gfire}
\bibinfo{author}{\bibfnamefont{K.~D.} \bibnamefont{Lozanov}} \bibnamefont{and}
  \bibinfo{author}{\bibfnamefont{M.~A.} \bibnamefont{Amin}},
  \bibinfo{journal}{Journal of Cosmology and Astroparticle Physics}
  \textbf{\bibinfo{volume}{2020}}, \bibinfo{pages}{058} (\bibinfo{year}{2020}).

\bibitem[{\citenamefont{Mancini et~al.}(2020)\citenamefont{Mancini, Del~Galdo,
  Chandramouli, Pagliai, and Barone}}]{mancini2020computational}
\bibinfo{author}{\bibfnamefont{G.}~\bibnamefont{Mancini}},
  \bibinfo{author}{\bibfnamefont{S.}~\bibnamefont{Del~Galdo}},
  \bibinfo{author}{\bibfnamefont{B.}~\bibnamefont{Chandramouli}},
  \bibinfo{author}{\bibfnamefont{M.}~\bibnamefont{Pagliai}}, \bibnamefont{and}
  \bibinfo{author}{\bibfnamefont{V.}~\bibnamefont{Barone}},
  \bibinfo{journal}{Journal of Chemical Theory and Computation}
  \textbf{\bibinfo{volume}{16}}, \bibinfo{pages}{5747} (\bibinfo{year}{2020}).

\bibitem[{\citenamefont{Manwadkar et~al.}(2020)\citenamefont{Manwadkar, Trani,
  and Leigh}}]{manwadkar2020chaos}
\bibinfo{author}{\bibfnamefont{V.}~\bibnamefont{Manwadkar}},
  \bibinfo{author}{\bibfnamefont{A.~A.} \bibnamefont{Trani}}, \bibnamefont{and}
  \bibinfo{author}{\bibfnamefont{N.~W.} \bibnamefont{Leigh}},
  \bibinfo{journal}{Monthly Notices of the Royal Astronomical Society}
  \textbf{\bibinfo{volume}{497}}, \bibinfo{pages}{3694} (\bibinfo{year}{2020}).

\bibitem[{\citenamefont{Ohno}(2020)}]{ohno2020}
\bibinfo{author}{\bibfnamefont{H.}~\bibnamefont{Ohno}}, \bibinfo{journal}{JOSA
  A} \textbf{\bibinfo{volume}{37}}, \bibinfo{pages}{411}
  (\bibinfo{year}{2020}).

\bibitem[{\citenamefont{Seki and Yunoki}(2021)}]{seki2021quantum}
\bibinfo{author}{\bibfnamefont{K.}~\bibnamefont{Seki}} \bibnamefont{and}
  \bibinfo{author}{\bibfnamefont{S.}~\bibnamefont{Yunoki}},
  \bibinfo{journal}{PRX Quantum} \textbf{\bibinfo{volume}{2}},
  \bibinfo{pages}{010333} (\bibinfo{year}{2021}).

\bibitem[{\citenamefont{Sheng et~al.}(2020)\citenamefont{Sheng, Shen, Tang,
  Wang, and Yuan}}]{sheng2020fast}
\bibinfo{author}{\bibfnamefont{C.}~\bibnamefont{Sheng}},
  \bibinfo{author}{\bibfnamefont{J.}~\bibnamefont{Shen}},
  \bibinfo{author}{\bibfnamefont{T.}~\bibnamefont{Tang}},
  \bibinfo{author}{\bibfnamefont{L.-L.} \bibnamefont{Wang}}, \bibnamefont{and}
  \bibinfo{author}{\bibfnamefont{H.}~\bibnamefont{Yuan}},
  \bibinfo{journal}{SIAM Journal on Numerical Analysis}
  \textbf{\bibinfo{volume}{58}}, \bibinfo{pages}{2435} (\bibinfo{year}{2020}).

\bibitem[{\citenamefont{Song et~al.}(2020)\citenamefont{Song, Vogt-Maranto,
  Wiscons, Matzger, and Tuckerman}}]{song2020generating}
\bibinfo{author}{\bibfnamefont{H.}~\bibnamefont{Song}},
  \bibinfo{author}{\bibfnamefont{L.}~\bibnamefont{Vogt-Maranto}},
  \bibinfo{author}{\bibfnamefont{R.}~\bibnamefont{Wiscons}},
  \bibinfo{author}{\bibfnamefont{A.~J.} \bibnamefont{Matzger}},
  \bibnamefont{and} \bibinfo{author}{\bibfnamefont{M.~E.}
  \bibnamefont{Tuckerman}}, \bibinfo{journal}{The Journal of Physical Chemistry
  Letters} \textbf{\bibinfo{volume}{11}}, \bibinfo{pages}{9751}
  (\bibinfo{year}{2020}).

\bibitem[{\citenamefont{Stern et~al.}(2021)\citenamefont{Stern, Alexahin,
  Burov, and Shiltsev}}]{stern2021self}
\bibinfo{author}{\bibfnamefont{E.}~\bibnamefont{Stern}},
  \bibinfo{author}{\bibfnamefont{Y.}~\bibnamefont{Alexahin}},
  \bibinfo{author}{\bibfnamefont{A.}~\bibnamefont{Burov}}, \bibnamefont{and}
  \bibinfo{author}{\bibfnamefont{V.}~\bibnamefont{Shiltsev}},
  \bibinfo{journal}{Journal of Instrumentation} \textbf{\bibinfo{volume}{16}},
  \bibinfo{pages}{P03045} (\bibinfo{year}{2021}).

\bibitem[{\citenamefont{Tong et~al.}(2021)\citenamefont{Tong, Xiong, He, Pan,
  and Zhu}}]{tong2021symplectic}
\bibinfo{author}{\bibfnamefont{Y.}~\bibnamefont{Tong}},
  \bibinfo{author}{\bibfnamefont{S.}~\bibnamefont{Xiong}},
  \bibinfo{author}{\bibfnamefont{X.}~\bibnamefont{He}},
  \bibinfo{author}{\bibfnamefont{G.}~\bibnamefont{Pan}}, \bibnamefont{and}
  \bibinfo{author}{\bibfnamefont{B.}~\bibnamefont{Zhu}},
  \bibinfo{journal}{Journal of Computational Physics} p.
  \bibinfo{pages}{110325} (\bibinfo{year}{2021}).

\bibitem[{\citenamefont{Wang et~al.}(2020{\natexlab{a}})\citenamefont{Wang,
  Nitadori, and Makino}}]{wang2020slow}
\bibinfo{author}{\bibfnamefont{L.}~\bibnamefont{Wang}},
  \bibinfo{author}{\bibfnamefont{K.}~\bibnamefont{Nitadori}}, \bibnamefont{and}
  \bibinfo{author}{\bibfnamefont{J.}~\bibnamefont{Makino}},
  \bibinfo{journal}{Monthly Notices of the Royal Astronomical Society}
  \textbf{\bibinfo{volume}{493}}, \bibinfo{pages}{3398}
  (\bibinfo{year}{2020}{\natexlab{a}}).

\bibitem[{\citenamefont{Wang et~al.}(2020{\natexlab{b}})\citenamefont{Wang, Fu,
  Zhang, and Zhao}}]{wang2020wave}
\bibinfo{author}{\bibfnamefont{Z.}~\bibnamefont{Wang}},
  \bibinfo{author}{\bibfnamefont{W.}~\bibnamefont{Fu}},
  \bibinfo{author}{\bibfnamefont{Y.}~\bibnamefont{Zhang}}, \bibnamefont{and}
  \bibinfo{author}{\bibfnamefont{H.}~\bibnamefont{Zhao}},
  \bibinfo{journal}{Physical Review Letters} \textbf{\bibinfo{volume}{124}},
  \bibinfo{pages}{186401} (\bibinfo{year}{2020}{\natexlab{b}}).

\bibitem[{\citenamefont{Wang et~al.}(2021{\natexlab{a}})\citenamefont{Wang,
  Sun, Liu, and Wu}}]{wang2021construction}
\bibinfo{author}{\bibfnamefont{Y.}~\bibnamefont{Wang}},
  \bibinfo{author}{\bibfnamefont{W.}~\bibnamefont{Sun}},
  \bibinfo{author}{\bibfnamefont{F.}~\bibnamefont{Liu}}, \bibnamefont{and}
  \bibinfo{author}{\bibfnamefont{X.}~\bibnamefont{Wu}}, \bibinfo{journal}{The
  Astrophysical Journal} \textbf{\bibinfo{volume}{907}}, \bibinfo{pages}{66}
  (\bibinfo{year}{2021}{\natexlab{a}}).

\bibitem[{\citenamefont{Wang et~al.}(2021{\natexlab{b}})\citenamefont{Wang,
  Sun, Liu, and Wu}}]{wang2021construction3}
\bibinfo{author}{\bibfnamefont{Y.}~\bibnamefont{Wang}},
  \bibinfo{author}{\bibfnamefont{W.}~\bibnamefont{Sun}},
  \bibinfo{author}{\bibfnamefont{F.}~\bibnamefont{Liu}}, \bibnamefont{and}
  \bibinfo{author}{\bibfnamefont{X.}~\bibnamefont{Wu}}, \bibinfo{journal}{arXiv
  preprint arXiv:2103.12272}  (\bibinfo{year}{2021}{\natexlab{b}}).

\bibitem[{\citenamefont{Wu et~al.}(2021)\citenamefont{Wu, Li, and
  Hou}}]{wu2021first}
\bibinfo{author}{\bibfnamefont{Y.}~\bibnamefont{Wu}},
  \bibinfo{author}{\bibfnamefont{H.}~\bibnamefont{Li}}, \bibnamefont{and}
  \bibinfo{author}{\bibfnamefont{J.}~\bibnamefont{Hou}},
  \bibinfo{journal}{Computational Materials Science}
  \textbf{\bibinfo{volume}{190}}, \bibinfo{pages}{110273}
  (\bibinfo{year}{2021}).

\bibitem[{\citenamefont{Xiao and Qin}(2021)}]{xiao2021explicit}
\bibinfo{author}{\bibfnamefont{J.}~\bibnamefont{Xiao}} \bibnamefont{and}
  \bibinfo{author}{\bibfnamefont{H.}~\bibnamefont{Qin}},
  \bibinfo{journal}{Plasma Science and Technology}  (\bibinfo{year}{2021}).

\bibitem[{\citenamefont{McLachlan}(1995)}]{mclachlan1995numerical}
\bibinfo{author}{\bibfnamefont{R.~I.} \bibnamefont{McLachlan}},
  \bibinfo{journal}{SIAM Journal on Scientific Computing}
  \textbf{\bibinfo{volume}{16}}, \bibinfo{pages}{151} (\bibinfo{year}{1995}).

\bibitem[{\citenamefont{Blanes and Moan}(2002)}]{blanes2002practical}
\bibinfo{author}{\bibfnamefont{S.}~\bibnamefont{Blanes}} \bibnamefont{and}
  \bibinfo{author}{\bibfnamefont{P.~C.} \bibnamefont{Moan}},
  \bibinfo{journal}{Journal of Computational and Applied Mathematics}
  \textbf{\bibinfo{volume}{142}}, \bibinfo{pages}{313} (\bibinfo{year}{2002}).

\bibitem[{\citenamefont{Fass{\`o}}(2003)}]{fasso2003comparison}
\bibinfo{author}{\bibfnamefont{F.}~\bibnamefont{Fass{\`o}}},
  \bibinfo{journal}{Journal of computational physics}
  \textbf{\bibinfo{volume}{189}}, \bibinfo{pages}{527} (\bibinfo{year}{2003}).

\bibitem[{\citenamefont{Higham}(2002)}]{higham2002accuracy}
\bibinfo{author}{\bibfnamefont{N.~J.} \bibnamefont{Higham}},
  \emph{\bibinfo{title}{Accuracy and stability of numerical algorithms}}
  (\bibinfo{publisher}{SIAM}, \bibinfo{year}{2002}).

\bibitem[{\citenamefont{Channell}(1986)}]{channell1986explicit}
\bibinfo{author}{\bibfnamefont{P.}~\bibnamefont{Channell}},
  \bibinfo{journal}{AT-6: ATN-86-6, Los Alamos National Laboratory}
  (\bibinfo{year}{1986}).

\bibitem[{\citenamefont{Hairer et~al.}(2008)\citenamefont{Hairer, McLachlan,
  and Razakarivony}}]{hairer2008achieving}
\bibinfo{author}{\bibfnamefont{E.}~\bibnamefont{Hairer}},
  \bibinfo{author}{\bibfnamefont{R.~I.} \bibnamefont{McLachlan}},
  \bibnamefont{and}
  \bibinfo{author}{\bibfnamefont{A.}~\bibnamefont{Razakarivony}},
  \bibinfo{journal}{BIT Numerical Mathematics} \textbf{\bibinfo{volume}{48}},
  \bibinfo{pages}{231} (\bibinfo{year}{2008}).

\bibitem[{\citenamefont{Antonana et~al.}(2018)\citenamefont{Antonana, Makazaga,
  and Murua}}]{antonana2018efficient}
\bibinfo{author}{\bibfnamefont{M.}~\bibnamefont{Antonana}},
  \bibinfo{author}{\bibfnamefont{J.}~\bibnamefont{Makazaga}}, \bibnamefont{and}
  \bibinfo{author}{\bibfnamefont{A.}~\bibnamefont{Murua}},
  \bibinfo{journal}{Numerical Algorithms} \textbf{\bibinfo{volume}{78}},
  \bibinfo{pages}{63} (\bibinfo{year}{2018}).

\bibitem[{\citenamefont{McLachlan et~al.}(2014)\citenamefont{McLachlan, Modin,
  and Verdier}}]{mclachlan2014symplectic}
\bibinfo{author}{\bibfnamefont{R.~I.} \bibnamefont{McLachlan}},
  \bibinfo{author}{\bibfnamefont{K.}~\bibnamefont{Modin}}, \bibnamefont{and}
  \bibinfo{author}{\bibfnamefont{O.}~\bibnamefont{Verdier}},
  \bibinfo{journal}{Physical Review E} \textbf{\bibinfo{volume}{89}},
  \bibinfo{pages}{061301} (\bibinfo{year}{2014}).

\bibitem[{\citenamefont{Rein and Spiegel}(2015)}]{rein2015ias15}
\bibinfo{author}{\bibfnamefont{H.}~\bibnamefont{Rein}} \bibnamefont{and}
  \bibinfo{author}{\bibfnamefont{D.~S.} \bibnamefont{Spiegel}},
  \bibinfo{journal}{Monthly Notices of the Royal Astronomical Society}
  \textbf{\bibinfo{volume}{446}}, \bibinfo{pages}{1424} (\bibinfo{year}{2015}).

\bibitem[{\citenamefont{Rodr{\'\i}guez and
  Barrio}(2012)}]{rodriguez2012reducing}
\bibinfo{author}{\bibfnamefont{M.}~\bibnamefont{Rodr{\'\i}guez}}
  \bibnamefont{and} \bibinfo{author}{\bibfnamefont{R.}~\bibnamefont{Barrio}},
  \bibinfo{journal}{Applied Numerical Mathematics}
  \textbf{\bibinfo{volume}{62}}, \bibinfo{pages}{1014} (\bibinfo{year}{2012}).

\bibitem[{\citenamefont{Seyrich and
  Lukes-Gerakopoulos}(2012)}]{seyrich2012symmetric}
\bibinfo{author}{\bibfnamefont{J.}~\bibnamefont{Seyrich}} \bibnamefont{and}
  \bibinfo{author}{\bibfnamefont{G.}~\bibnamefont{Lukes-Gerakopoulos}},
  \bibinfo{journal}{Physical Review D} \textbf{\bibinfo{volume}{86}},
  \bibinfo{pages}{124013} (\bibinfo{year}{2012}).

\bibitem[{\citenamefont{Symes et~al.}(2016)\citenamefont{Symes, McLachlan, and
  Blakie}}]{symes2016efficient}
\bibinfo{author}{\bibfnamefont{L.}~\bibnamefont{Symes}},
  \bibinfo{author}{\bibfnamefont{R.}~\bibnamefont{McLachlan}},
  \bibnamefont{and} \bibinfo{author}{\bibfnamefont{P.}~\bibnamefont{Blakie}},
  \bibinfo{journal}{Physical Review E} \textbf{\bibinfo{volume}{93}},
  \bibinfo{pages}{053309} (\bibinfo{year}{2016}).

\bibitem[{\citenamefont{Vilmart}(2008)}]{vilmart2008reducing}
\bibinfo{author}{\bibfnamefont{G.}~\bibnamefont{Vilmart}},
  \bibinfo{journal}{Journal of computational physics}
  \textbf{\bibinfo{volume}{227}}, \bibinfo{pages}{7083} (\bibinfo{year}{2008}).

\bibitem[{\citenamefont{Blanes et~al.}(2006)\citenamefont{Blanes, Casas, and
  Murua}}]{blanes2006composition}
\bibinfo{author}{\bibfnamefont{S.}~\bibnamefont{Blanes}},
  \bibinfo{author}{\bibfnamefont{F.}~\bibnamefont{Casas}}, \bibnamefont{and}
  \bibinfo{author}{\bibfnamefont{A.}~\bibnamefont{Murua}},
  \bibinfo{journal}{SIAM Journal on Scientific Computing}
  \textbf{\bibinfo{volume}{27}}, \bibinfo{pages}{1817} (\bibinfo{year}{2006}).

\bibitem[{\citenamefont{Wisdom et~al.}(1996)\citenamefont{Wisdom, Holman, and
  Touma}}]{wisdom1996symplectic}
\bibinfo{author}{\bibfnamefont{J.}~\bibnamefont{Wisdom}},
  \bibinfo{author}{\bibfnamefont{M.}~\bibnamefont{Holman}}, \bibnamefont{and}
  \bibinfo{author}{\bibfnamefont{J.}~\bibnamefont{Touma}},
  \bibinfo{journal}{Fields Institute Communications}
  \textbf{\bibinfo{volume}{10}}, \bibinfo{pages}{217} (\bibinfo{year}{1996}).

\end{thebibliography}

\end{document}